\documentclass[a4paper]{jpconf}
\usepackage{graphicx,amssymb}
\begin{document}
\title{Violation of the Carter-Israel conjecture and its astrophysical implications}

\author{Cosimo Bambi}

\address{Institute for the Physics and Mathematics of the Universe, The University of Tokyo \\ 5-1-5 Kashiwanoha, Kashiwa-shi, Chiba 277-8583, Japan}

\ead{cosimo.bambi@ipmu.jp}

\begin{abstract}
On the basis of the Carter-Israel conjecture, today we believe that some compact and massive objects in the Galaxy and in the Universe are Kerr black holes. However, this idea cannot yet be confirmed by observations. We can currently obtain reliable estimates of the masses of these objects, but we do not know if the space-time around them is described by the Kerr metric and if they have an event horizon. A fundamental limit for a Kerr black hole is the Kerr bound $|a_*| \le 1$. Here I discuss some astrophysical implications associated with the violation of this bound, which can thus be used to test the Carter-Israel conjecture.
\end{abstract}

\section{Introduction}

Today we believe that the final product of the gravitational collapse is a Kerr-Newman black hole. This is the Carter-Israel conjecture and it is based on the following argument. First, there are some singularity theorems showing that in general relativity the collapsing matter produces space-time singularities~\cite{hawking}. These theorems do not say anything about the nature of the singularities. The second step is thus to assume the Weak Cosmic Censorship conjecture~\cite{penrose}, according to which space-time singularities must be hidden behind an event horizon. Surprisingly, in general relativity in 4 dimensions, the only possibility is the Kerr-Newman space-time~\cite{no-hair1,no-hair2,no-hair3,no-hair4}.

The Kerr-Newman metric has three parameters: the mass of the object, $M$, its spin, $J$, and its electric charge, $Q$. The spin $J$ can be replaced by the spin parameter $a = J/M$, or by the dimensionless spin parameter $a_* = J/M^2$. For macroscopic bodies, the electric charge is usually very small and can be neglected\footnote{For example, the electric charge is important for black holes with a mass $M \lesssim 10^{20}$~g in a ionized plasma~\cite{dolgov}. Such small black holes cannot be produced today in the Universe, but could have been produced in the early Universe, when the matter density was much higher.}. Hereafter, we will thus restrict our attention to Kerr space-times with $Q = 0$. A fundamental property of Kerr black holes is the Kerr bound $|a_*| \le 1$, which is the condition for the existence of the event horizon.

Observational evidences supporting the Carter-Israel conjecture are still definitively weak~\cite{narayan}. Astronomical observations have led to the discovery of at least two classes of astrophysical black hole candidates\footnote{The existence of a third class of astrophysical black holes, intermediate mass objects with $M \sim 10^2 - 10^4$~$M_\odot$, is still quite controversial, because there are no reliable dynamical measurements of their masses.}: stellar-mass objects in X-ray binary systems ($M \sim 5 - 20$~$M_\odot$) and super-massive objects at the center of many galaxies ($M \sim 10^5 - 10^{10}$~$M_\odot$). All these objects are supposed to be Kerr black holes because they cannot be explained with something else without introducing new physics. For example, stellar-mass black hole candidates in X-ray binary systems are too heavy to be neutron/quark stars for any reasonable matter equation of state~\cite{bh1a,bh1b}. The super-massive black hole candidate at the center of the Galaxy is too massive and compact to be a cluster of non-luminous bodies~\cite{bh2}. On the other hand, we do not have any observational evidence of the Kerr metric and of the existence of the event horizon.

New physics may instead be expected: it is indeed difficult to believe that space-time singularities can exist, even if behind an event horizon. So, the assumption of the Weak Cosmic Censorship conjecture seems to be a quite artificial trick, to exclude space-times in which new physics can be causally connected to distant observers. In other words, the conjecture would be motivated by our poor knowledge of the theory at high energy, but deviations from the Kerr space-times may be possible in Nature. Motivated by such a simple and general argument, one can consider the possibility that the Carter-Israel conjecture can be violated. Here, I will discuss the case of super-spinars, compact objects with $|a_*| > 1$. I will review the basic features of the accretion process onto super-spinars; more details on the subject can be found in the original papers~\cite{sim1, sim2, sim3}. Other observational properties of super-spinars were studied in~\cite{ss1,ss2,ss3}.

\section{Super-spinars}

The Weak Cosmic Censorship conjecture was originally motivated by the fact that space-times with naked singularities present several kind of pathologies. One can however interpret such pathological features with the necessity of new physics. For example, there is a connection between the existence of naked singularities and regions with closed time-like curves~\cite{nk-ctc}. The physical interpretation of space-times with causality violating regions has been recently investigated by some authors in Supergravity~\cite{sol1,sol2,sol3,sol4}. Here, one finds space-times which apparently cannot be ruled out as unphysical and where causality can be violated. The solution of this puzzle seems to be in high energy corrections of the theory: at least in some cases, one can expect that the space-time goes to a new phase and a domain wall forms. Across the domain wall, the metric is non-differentiable and the expected region with closed time-like curves arises from the naive continuation of the metric ignoring the domain wall. The latter can be made of very exotic stuff, like super-tubes~\cite{sol1,sol3,sol4} or fundamental strings~\cite{sol2}. It is also remarkable that we know several counterexamples which look physically reasonable and in which the collapsing matter starts from regular initial data and evolves into a naked singularity, see e.g. Refs.~\cite{cex1,cex2,cex3,cex4,cex5,cex6,cex7}.

The simplest object violating the Carter-Israel conjecture is probably the super-spinar~\cite{g-h}, a compact body with $|a_*| > 1$. In absence of a uniqueness theorem similar to the one for Kerr black holes in 4 dimensions, super-spinars may be quite complex objects, characterized by many parameters. Nevertheless, at first approximation one can still expect to be able to describe their gravitational field with the Kerr metric. The first two terms in a multipole moment expansion of the space-time correspond to the mass and the spin of the massive object, while its deformation is encoded in higher order moments, which are typically much less important. In other words, deviations from the Kerr metric are usually very small, as one can see in~\cite{tomi-sato}. If a compact object has $|a_*| > 1$, it cannot be a Kerr black hole. So, the Kerr bound can be used to test the Carter-Israel conjecture.

In absence of a complete theory of gravity, we have to take a phenomenological approach to study the astrophysical properties of super-spinars. They cannot be extremely compact, like a naked singularity, because otherwise they would be unstable due to the ergoregion instability~\cite{enrico}. It is probably reasonable to take their radius of order of their gravitational radius, which makes them more similar to a relativistic star made of very exotic matter than to a naked singularity.

Another issue is how super-spinars can be created. At present, it is not clear if it is possible to overspin an existent black hole~\cite{j-s}. If super-spinars can be the final product of the gravitational collapse of a star, then one should be able to obtain a super-spinning object from numerical simulations of a collapsing star. However this is not an easy job and there are so many unknown ingredients (and unknown physics) that it is unlikely to get an answer in a near future. We can however notice that recent attempts to measure the spin of stellar-mass black hole candidates suggest that these objects can rotate very fast~\cite{mcclintock}\footnote{Let us notice that current estimates of black hole spin assume $|a_*| \le 1$ and cannot be used to say that these objects are not super-spinars. If we allow for any value of the spin parameter, we would obtain two estimates, one with $|a_*| \le 1$ and another with $|a_*| > 1$, because of the degeneracy discussed in~\cite{ss3}.}. Even if these measurements should be taken with great caution, it is intriguing the case of GSR~$1915+105$, whose spin parameter is estimated to be in the range $0.98 - 1$. Since the evolution of the spin parameter due to accretion should be negligible in this system, the estimated $a_*$ would reflect the initial spin of the object after its formation. One can thus think that the gravitational collapse of a star can produce a very fast rotating object. Since from the theory of stellar evolution we expect around $10^8$ stellar-mass black holes in the Galaxy, even a low probability of violating the bound $|a_*| \le 1$ may lead to a population of super-spinars in the Galaxy.

\section{Numerical study of the accretion process}

The study of the accretion process plays a fundamental role in the physics of compact objects, because it is the accretion process that determines how radiation is released by the accreting matter, and so what we can see from the compact object. It is thus not surprising that the accretion process onto Kerr black holes has been well studied in the literature and many research groups work on the subject~\cite{chak,font}. The quasi-steady-state configuration of adiabatic and spherically symmetric accretion onto a Schwarzschild black hole can be studied analytically~\cite{michel}. However, in general, a numerical approach is necessary. The first numerical hydrodynamics simulations of the accretion process onto black holes can be traced back to the work of Wilson in 1972~\cite{wilson}, and were then extended in~\cite{hsw1,hsw2}. After these works, the research was mainly devoted to the study of accretion from thick disks and tori, and to the study of the tori instabilities~\cite{h91,yokosawa,i-b,devilliers}.

In~\cite{sim1,sim2,sim3}, I studied numerically the accretion process in Kerr space-time with arbitrary value of the spin parameter $a_*$. I neglected the back-reaction of the fluid to the geometry of the space-time, as well as the increase in mass and the variation in spin of the central object due to accretion. Such an approximation is surely reasonable if we want to study a stellar-mass compact object in a binary system, because in this case the matter captured from the stellar companion is typically small in comparison with the total mass of the compact object. The results of this simulations should not be applied to long-term accretion onto a super-massive object at the center of a galaxy, where accretion makes the mass of the compact object increase by a few orders of magnitude from its original value. The accreting matter was modeled as a perfect fluid.

The master formulas are the equations of conservation of baryon number and of the fluid energy-momentum tensor
\begin{eqnarray}\label{eq-cons}
\nabla_\mu J^\mu = 0 \, , \quad
\nabla_\mu T^{\mu\nu} = 0 \, .
\end{eqnarray}
Here $J^\mu = \rho u^\mu$ and $T^{\mu\nu} = \rho h u^\mu u^\nu + p g^{\mu\nu}$, where $\rho$ is the rest-mass energy density, $p$ is the pressure, $u^\mu$ is the fluid 4-velocity, $h = 1 + \epsilon + p/\rho$ is the specific enthalpy, and $\epsilon$ is the specific internal energy density. In order to solve the system, an equation of state $p = p (\rho,\epsilon)$ must be specified.

The calculations were made with the relativistic hydrodynamics module of the public available code PLUTO~\cite{pluto1,pluto2}, properly modified for the case of curved space-time. I used the $3+1$ Eulerian formalism of Ibanez and collaborators~\cite{ibanez}. The line element of the space-time is written in the form
\begin{eqnarray}
ds^2 = - \left(\alpha^2 - \beta_i \beta^i\right) dt^2 
+ 2 \beta_i dt dx^i + \gamma_{ij} dx^i dx^j \, ,
\end{eqnarray}
where $\alpha$ is the lapse function, $\beta^i$ is the shift vector, and $\gamma_{ij}$ is the 3-metric induced on each space-like slice. Here it is convenient to use two set of variables. The {\it primitive variables} are
\begin{eqnarray}
{\bf V} = \left( \rho, v^i, p \right)^T
\end{eqnarray}
and are the quantities whose evolution and quasi-steady-state (if any) we want to determine. The hydrodynamical equations are instead solved in term of the {\it conserved variables}
\begin{eqnarray}
{\bf U} = \left( D, S_i, \tau \right)^T \, ,
\end{eqnarray}
which can be written in term of the primitive ones as $D = \rho W$, $S_i = \rho h W^2 v_i$, and $\tau = \rho h W^2 - D - p$, where $W$ is the Lorentz factor. The equations of conservation (\ref{eq-cons}) can now be written as
\begin{eqnarray}
\frac{1}{\sqrt{-g}} \left[
\frac{\partial}{\partial t} 
\left( \sqrt{\gamma} \, {\bf U} \right)
+ \frac{\partial}{\partial x^i} 
\left( \sqrt{-g} \, {\bf F}^i \right) 
\right] = {\bf \mathcal S} \, ,
\end{eqnarray}
where ${\bf F}^i$ and ${\bf \mathcal S}$ are defined by
\begin{eqnarray}
{\bf F}^i &=& \Big( 
D \left( v^i - \beta^i/\alpha \right) , \;
S_j \left(v^i - \beta^i/\alpha \right) + p \delta_j^i , \;
\tau \left(v^i - \beta^i/\alpha \right) + p v^i \; 
\Big)^T \, , \\
{\bf \mathcal S} &=& \Big( 
0 , \;
T^{\mu\nu} \left( \partial_\mu g_{\nu j} 
- \Gamma^{\lambda}_{\mu\nu} g_{\lambda j} \right) , \;
\alpha \left( T^{\mu 0} \partial_\mu \ln\alpha 
- T^{\mu\nu} \Gamma^{0}_{\mu\nu} \right) \; 
\Big)^T \, .
\end{eqnarray}

\section{Results} 

Roughly speaking, naked singularities are not hidden behind an event horizon because their gravitational field is too weak to trap light rays. Close to the expected naked singularity, the gravitational field may be even repulsive~\cite{rep1,rep2,rep3,rep4}. So, a quasi-steady-state accretion flow onto a naked singularity may be impossible: in some cases, the gas is accumulated around the massive object, forming a high density cloud that continues to grow~\cite{sim1, babichev}.

In~\cite{sim1,sim2,sim3}, I considered adiabatic and spherically symmetric (at large radii) accretion process onto Kerr black holes and Kerr super-spinars. The initial configuration is a static cloud of gas around the massive object and then the system evolves to find a quasi-steady-state configuration. Gas is injected into the computational domain from the outer boundary at a constant rate and isotropically. With this set-up, the two parameters determining the accretion process are the spin parameter, $a_*$, and the radius of the compact object, $R$.

\begin{figure}[ht]
\includegraphics[width=17pc]{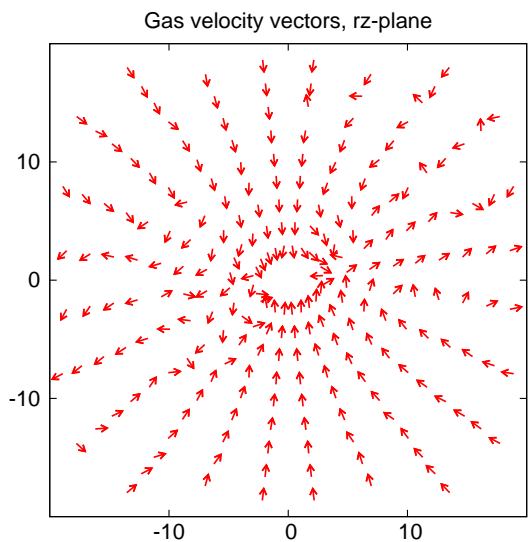}
\hspace{2pc}
\begin{minipage}[b]{17pc}
\caption{\label{f0} Snapshot of the direction of the gas velocity around a super-spinar with $a_* = 3.0$, on a plane containing the axis of symmetry $z$. Here matter accretes from the poles, while the repulsive gravitational field produces outflows around the equatorial plane. The unit of length along the axes is $M$.}
\end{minipage}
\end{figure}

\begin{figure}[ht]
\begin{minipage}{17pc}
\includegraphics[width=17pc]{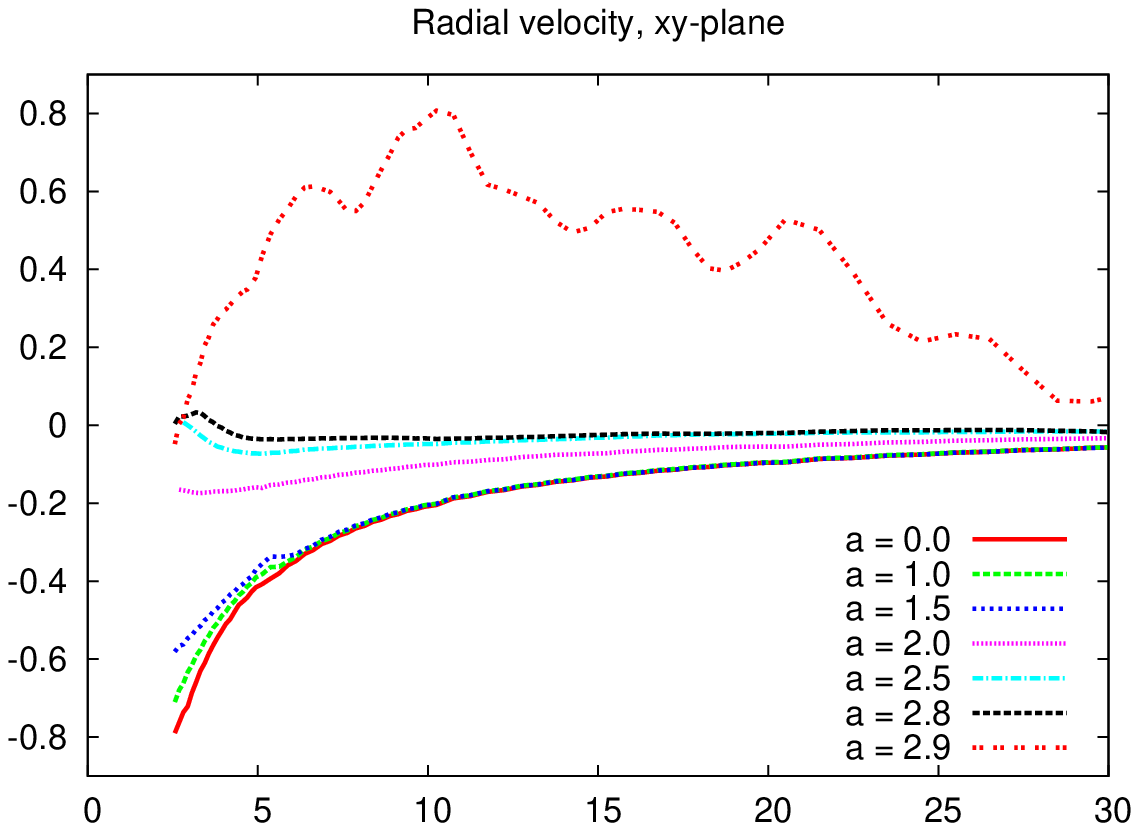}
\caption{\label{f1} Snapshot at $t = 500$~$M$ of the radial velocity as a function of the radial coordinate on the equatorial plane. $R = 2.5$~$M$, radial coordinate in units $M=1$.}
\end{minipage}
\hspace{2pc}
\begin{minipage}{17pc}
\includegraphics[width=17pc]{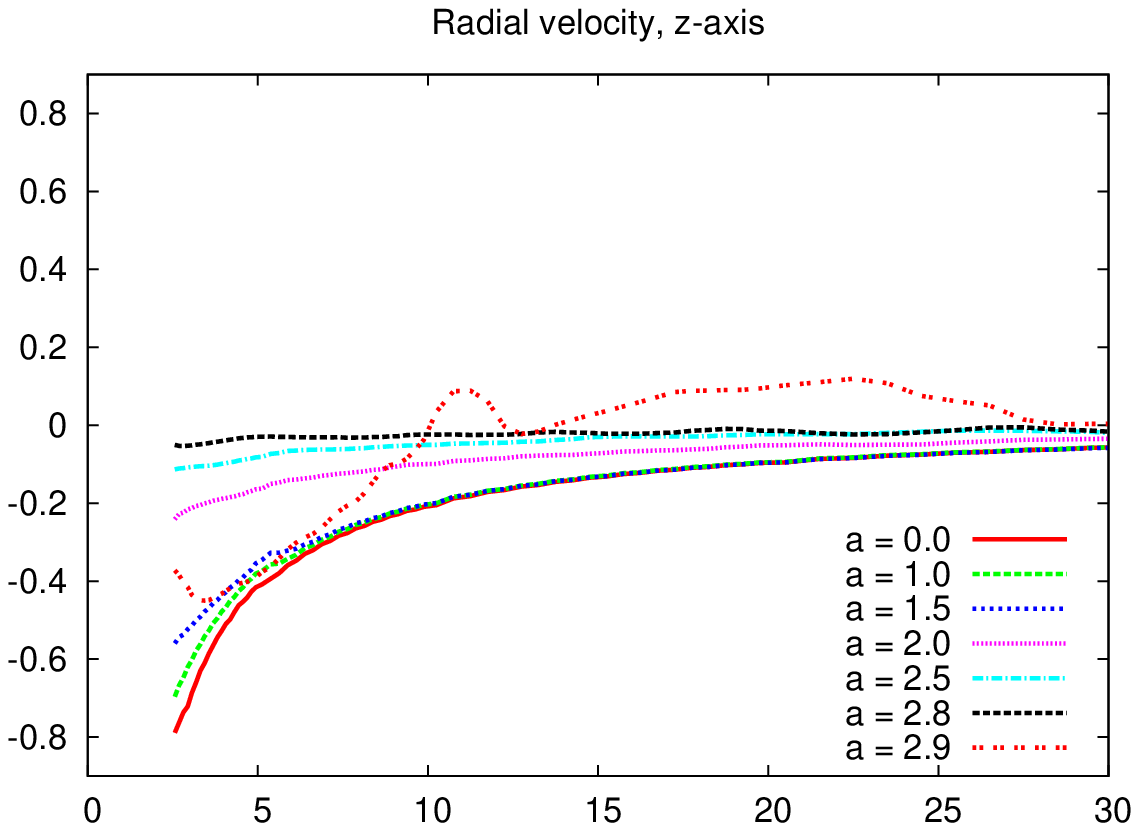}
\caption{\label{f2} Snapshot at $t = 500$~$M$ of the radial velocity as a function of the radial coordinate along the $z$-axis. $R = 2.5$~$M$, radial coordinate in units $M=1$.}
\end{minipage} 
\end{figure}

\begin{figure}[ht]
\begin{minipage}{17pc}
\includegraphics[width=17pc]{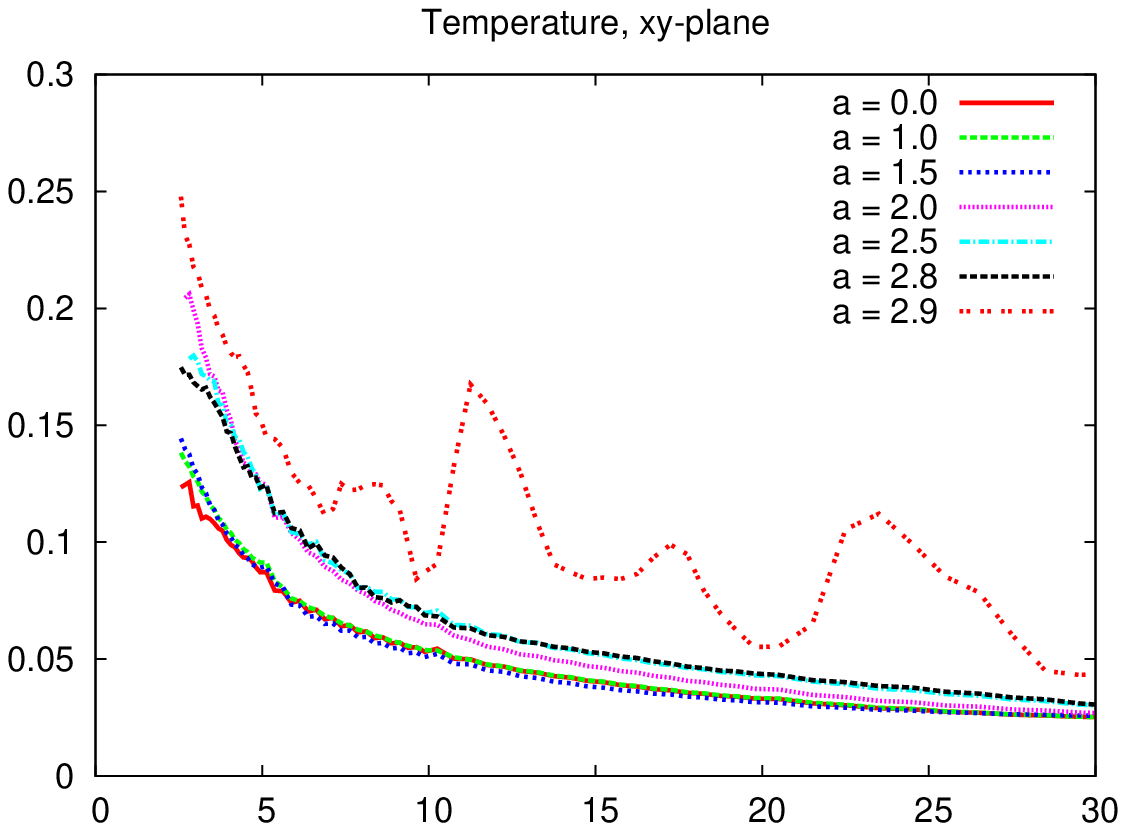}
\caption{\label{f3} Snapshot at $t = 500$~$M$ of the temperature as a function of the radial coordinate on the equatorial plane. $R = 2.5$~$M$, temperature in GeV, radial coordinate in units $M=1$.}
\end{minipage}
\hspace{2pc}
\begin{minipage}{17pc}
\includegraphics[width=17pc]{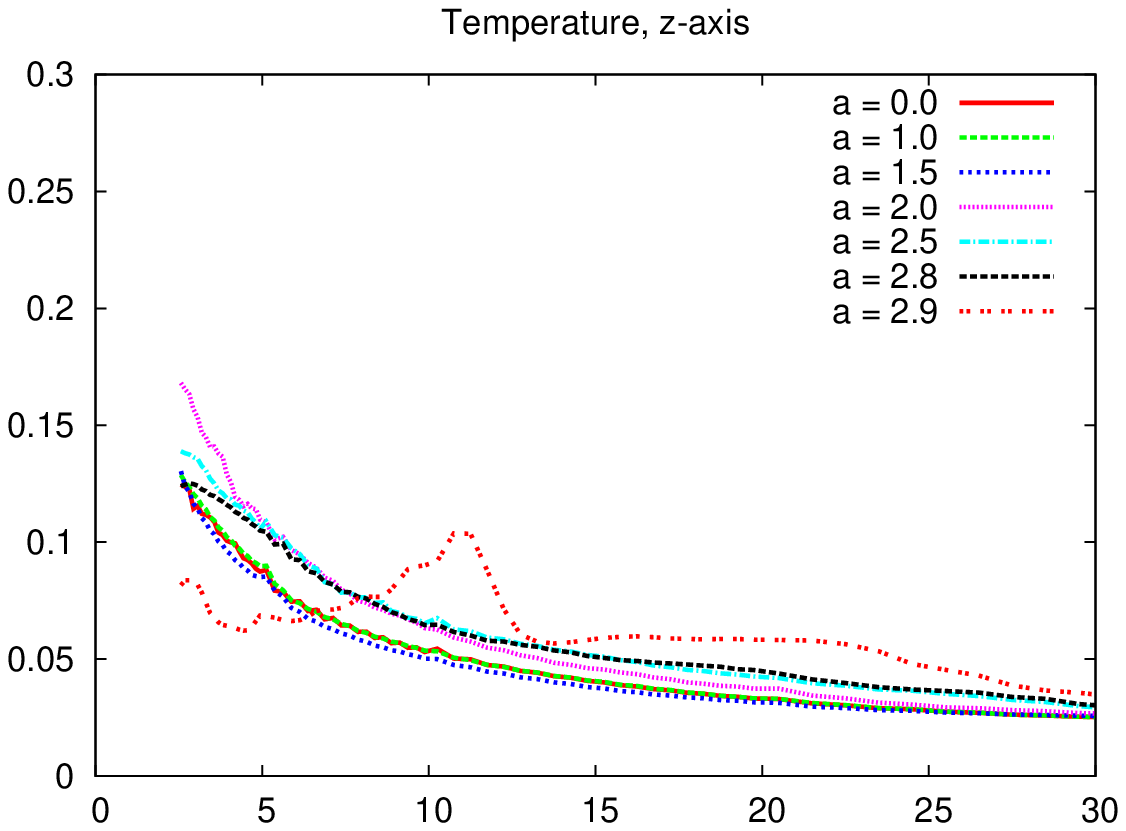}
\caption{\label{f4} Snapshot at $t = 500$~$M$ of the temperature as a function of the radial coordinate along the $z$-axis. $R = 2.5$~$M$, temperature in GeV, radial coordinate in units $M=1$.}
\end{minipage} 
\end{figure}

One can distinguish three kinds of accretion~\cite{sim2,sim3}:
\begin{enumerate}
\item {\it Black hole accretion}. For black holes and super-spinars with $|a_*|$ moderately larger than 1, one finds the usual accretion process onto a compact object. For a given $R$, the increase in $|a_*|$ makes the accretion process more difficult: in the quasi-steady-state configuration, the velocity of the gas around the compact object is lower, while the density and the temperature are higher. The gravitational field indeed 
becomes weaker for higher spin parameter, as one can easily understand by noticing that the radius of the event horizon of a black hole monotonically decreases with $a_*$. The difference, however, is very small and the exact value of the spin parameter does not affect significantly the process. 
\item {\it Intermediate accretion}. As the spin parameter increases, the gravitational force around the super-spinar becomes weaker and even repulsive. Now the accretion process is significantly suppressed: the flow around the super-spinar becomes subsonic and the density and the temperature of the gas increase further. 
\item {\it Super-spinar accretion}. For high value of the spin parameter, the process of accretion is very different: matter accretes from the poles, while the repulsive gravitational field produces outflows around the equatorial plane (see Fig.~\ref{f0}). 3-dimensional simulations show that the production of outflows is a quite chaotic phenomenon, without the formation of a stable structure~\cite{sim3}. 
\end{enumerate}  
In Figs.~\ref{f1} and \ref{f2}, I have plotted the quantity $v^{(r)} = e^{(r)}_{i} v^i$, where $e^{(a)}_{b}$ is the tetrad of a locally non-rotating observer~\cite{bpt}. Figs.~\ref{f3} and \ref{f4} show the temperature profile around black holes and super-spinars. In these simulations, $R = 2.5$~$M$ and the gas has a polytropic index $\Gamma = 5/3$ (non-relativistic particles). However, the qualitative behavior of the accretion process does not depend on the gas equation of state. For $a_* = 0$, 1, and 1.5, we find a black hole-like accretion; for $a_* = 2$, an intermediate accretion; for $a_* = 2.9$, the accretion is of super-spinar type. When $a_* = 2.5$ and 2.8, the accretion is essentially of the second kind, but there is some very weak ejection of matter near the equatorial plane.

\begin{figure}[ht]
\includegraphics[width=17pc]{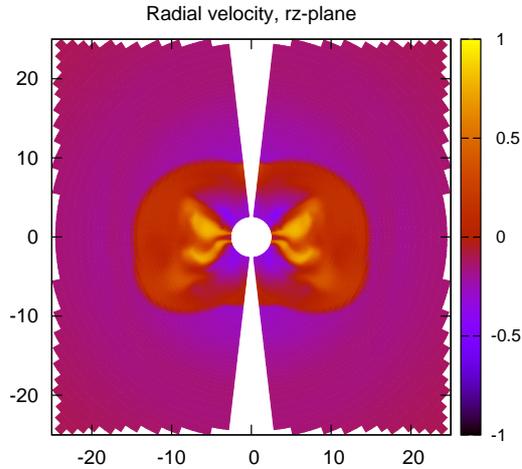}
\hspace{2pc}
\begin{minipage}[b]{17pc}
\caption{\label{f5} Radial velocity of the gas around a super-spinar with $a_* = 2.8$ and $R = 2.5$~$M$.}
\end{minipage}
\end{figure}

Unlike jets and outflows produced around black holes, here the outflows are produced by the repulsive gravitational force at a small distance from the super-spinar and are ejected around the equatorial plane. These outflows become more energetic for higher value of the spin parameter. In some circumstances, the amount of matter in the outflow is considerable, which can indeed significantly reduce the mass accretion rate. On the other hand, for lower values of the spin parameter, the outflows may not be energetic enough to be ejected at large radii and escape from the gravitational field of the object. In these cases, one finds a convective region around the super-spinar, where the ejected gas is pushed back by the accreting fluid. This possibility is shown in Fig.~\ref{f5}, for the case of a super-spinar with $a_* = 2.8$ and a gas made of relativistic particles ($\Gamma = 4/3$).

Since the repulsive gravitational force around super-spinars seems to be able to create collimated jets with high Lorentz factor, in~\cite{sim2} I put forward the possibility that long gamma ray bursts might be explained with the formation of a super-spinar.

\ack

I would like to thank Naoki Yoshida for careful reading of the first draft of this manuscript. This work was supported by World Premier International Research Center Initiative (WPI Initiative), MEXT, Japan, and by the JSPS Grant-in-Aid for Young Scientists (B) No. 22740147.

\end{document}